\documentclass[aps,prb,reprint,superscriptaddress,twocolumn,showkeys,amsmath,amssymb,longbibliography]{revtex4-2}
\usepackage[english]{babel}
\usepackage{amsmath,amssymb,bbm,mathrsfs,bm,braket,color,graphicx,comment,amsfonts,dsfont}
\usepackage[colorlinks,citecolor=blue,urlcolor=blue]{hyperref}
\usepackage[mathscr]{euscript}
\usepackage[normalem]{ulem}
\usepackage{comment}
\newcommand{\pd}{{\phantom{\dag}}}

\begin{document}
\title{Mixed higher-order topology: boundary non-Hermitian skin effect induced by a Floquet bulk}
\author{Hui Liu}
\affiliation{IFW Dresden and W{\"u}rzburg-Dresden Cluster of Excellence ct.qmat, Helmholtzstrasse 20, 01069 Dresden, Germany}
\author{Ion Cosma Fulga}
\affiliation{IFW Dresden and W{\"u}rzburg-Dresden Cluster of Excellence ct.qmat, Helmholtzstrasse 20, 01069 Dresden, Germany}

\begin{abstract}
We show that anomalous Floquet topological insulators generate intrinsic, non-Hermitian topology on their boundary.
As a consequence, removing a boundary hopping from the time-evolution operator stops the propagation of chiral edge modes, leading to a non-Hermitian skin effect.
This does not occur in Floquet Chern insulators, however, in which boundary modes continue propagating.
The non-Hermitian skin effect on the boundary is a consequence of the nontrivial topology of the bulk Floquet operator, which we show by designing a real-space topological invariant.
Our work introduces a form of `mixed' higher-order topology, providing a bridge between research on periodically-driven systems and the study of non-Hermiticity.
It suggests that periodic driving, which has already been demonstrated in a wide range of experiments, may be used to generate non-Hermitian skin effects.
\end{abstract}

\maketitle

\section{Introduction}
\label{sec:introduction}
The Hermiticity of Hamiltonians is a principal feature of quantum physics.
As a consequence, time-evolution operators are unitary, with eigenvalues constrained to be phase factors, thus maintaining probability conservation. 
In periodically-driven systems, commonly characterized in terms of their Floquet operators, this means that quasienergies $\varepsilon$ are only defined modulo $2\pi/T$, with $T$ being the driving period.

The periodicity of Floquet eigenphases can have important consequences in the context of band topology~\cite{Hasan_rmp, Qi_rmp, Chiu_rmp}, 
since it provides an extra bulk quasienergy gap (at $\varepsilon=\pm \pi/T$) to topological boundary modes.
When this gap is nontrivial, the resulting Floquet topological insulators are completely induced by the time-periodic driving, and thus have no counterpart in static systems \cite{Lindner_np, Oka_prb, Inoue_prl, Kitagawa_prb, Lindner_prb, Torres_prb, Katan_prl, Qin_floquet}. 
An example of such a phase is the so-called anomalous Floquet topological insulator (AFTI)~\cite{rlbl_prx}, a two-dimensional (2D) system in which each gap contains the same, nonzero number of chiral edge states, despite all bulk bands being topologically trivial.
As a result, the chiral edge modes wind around the $[-\pi/T, \pi/T)$ quasienergy zone, or equivalently, around the unit circle in the complex plane.

When Hermiticity is broken, the emergence of complex Hamiltonian eigenvalues can alter the well-established, fundamental concepts of band topology and yield new phenomena~\cite{Okuma_review, Emil_review, Ashida_review}. 
An interesting example is the non-Hermitian skin effect, in which an extensive number of modes pile up at the boundaries of a system~\cite{Yao_prl, Lee_prl, Kunst_prl, Xiong_2018, Yao_chern, Yokomizo_prl, Zhang_prl, Longhi_prl, Borgnia_prl, Zhang_nc, Li2020, Martinez_nh, Qin_nhermitian, Ghosh_prb}.
One of the simplest models showing this behavior is the Hatano-Nelson model \cite{HN_prl}, a one-dimensional chain with nearest-neighbor hoppings that are nonreciprocal.
The skin modes that occur in this system when open boundary conditions (OBCs) are imposed have a topological origin. They are protected by the winding number of the infinite system spectrum~\cite{topological_origin}.

The topology of unitary operators is mainly studied in a Hermitian context, with a primary focus on periodically-driven, Hermitian Hamiltonians \cite{Nathan_2015, Roy_prb_fti, Harper_review, Yao_prb, Else_prb, zhang_ruixing, Oka_review, Fulga_fti}.
Here we examine unitary operators from the point of view of non-Hermiticity, showing that this leads to an alternate form of higher-order topology.
Intuitively, both the Hatano-Nelson model and the AFTI are characterized by a nonzero spectral winding, of the bulk states in the first case and of the chiral edge modes in the second case.
We show that these two windings are in fact connected: the bulk Floquet topology induces the formation of a non-Hermitian topological chain at the system boundary, without the need for any additional perturbation.

Working with one of the most well-known AFTI models \cite{rlbl_prx}, we found that removing a hopping from the boundary of the Floquet operator stops the propagation of chiral edge modes; they pile up at the defect position instead.
This is not a property of all Floquet chiral edge modes, however, but of those which have a nonzero spectral winding.
When chiral edge modes exist without spectral winding, such as in a Floquet Chern insulator phase, the skin effect is not robust, and the edge modes can continue propagating around the defect.
These two different behaviors can be predicted by a real-space invariant computed directly from the full, 2D Floquet operator, confirming the presence of a mixed higher-order topology.

The rest of the work is organized as follows. 
In Sec. II, we introduce the Floquet system and relate it to non-Hermitian topology. 
In Sec. III, we cut the edge to show the boundary skin effect in an two-dimensional anomalous Floquet topological phase. 
The topological protection of this phenomenon is studied in Sec. IV. 
We conclude in Sec. V.

\section{Non-Hermitian topology in a Floquet system}
\label{sec:model}
\begin{figure}
	\centering
	\includegraphics[width=\linewidth]{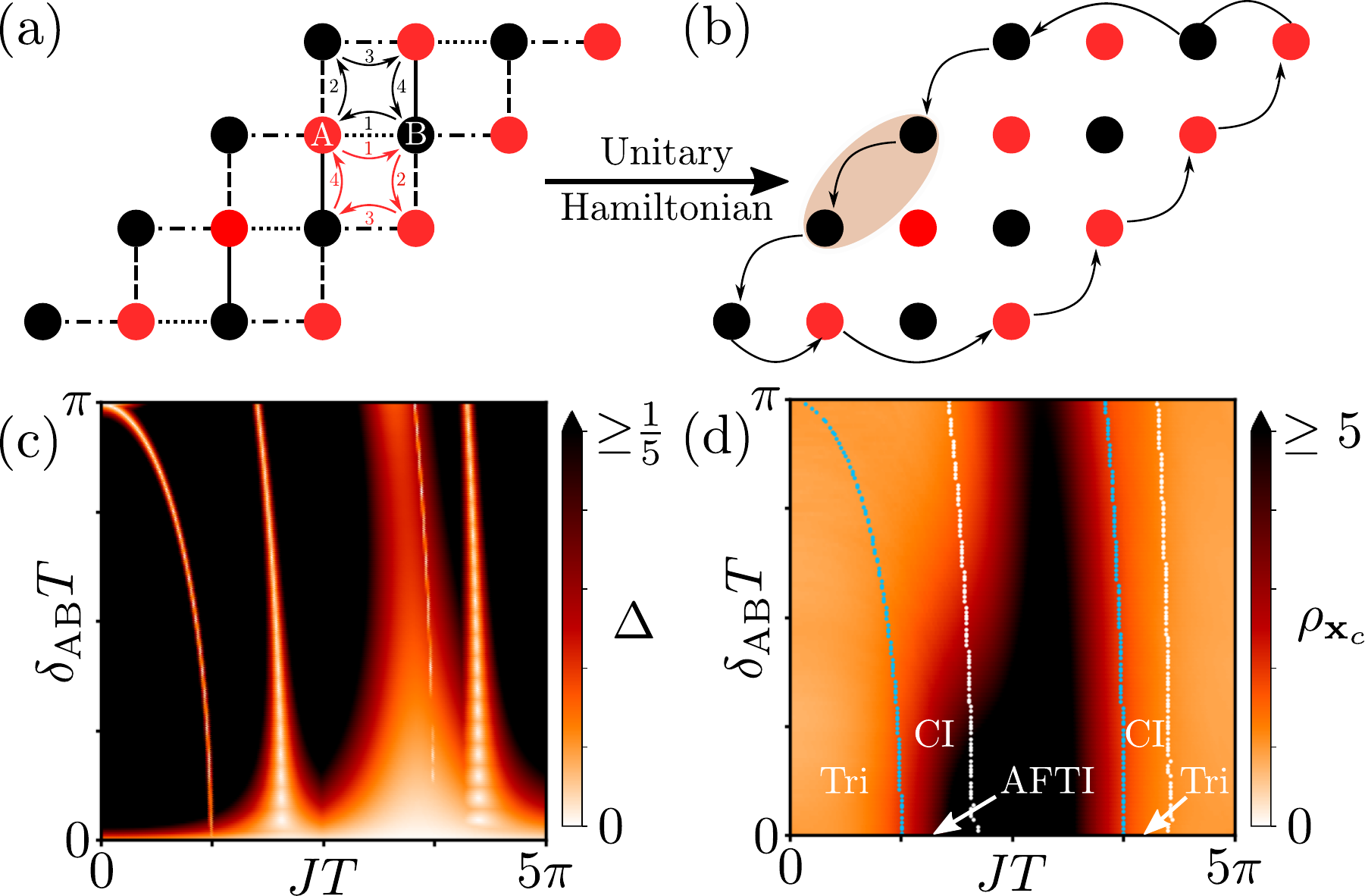}
	\caption{(a) Periodically-driven Hamiltonian model, with red and black dots representing the sublattices A and B.
	The dotted, dashed, dot-dashed, and solid lines correspond to the hopping terms of the time step $1$, $2$, $3$, and $4$, respectively. 
	The red (black) arrows indicate the evolutions of bulk modes on sublattice A (B) at resonant driving.
	(b) Real-space Floquet operator at resonant driving. 
	The black arrows denote unidirectional hoppings, identical to those of the Hatano-Nelson model in the limit of maximal nonreciprocity. 
	The edge cutting corresponds to removing the hopping from the shaded area. 
	(c) The smallest of the two bulk gaps (at $\varepsilon=0$ and $\pi/T$), $\Delta=\text{min}(\Delta_{0},~\Delta_{\pi/T})$, is plotted as a function of $J$ and $\delta_{\text{AB}}$.
	(d) Map of $\rho_{{\bf x}_c}$, the real-space probability density summed over all states, corresponding to the site adjacent to the cut hopping [top site enclosed in the ellipse of panel (b)].
	Larger densities (darker colors) suggest the appearance of a non-Hermitian skin effect.
	The white and blue dotted lines show the approximate location of gap closings at $\varepsilon=0$ and $\pi/T$.
	The trivial (Tri), Chern insulator (CI), and AFTI phases are indicated.
	White arrows show which phases replace the CIs at $\delta_{\text{AB}}=0$.
	See SM for numerical details \cite{SupMat}.
		\label{fig:rlbl_model} 
	}
\end{figure}

We start with the Rudner-Lindner-Berg-Levin model~\cite{rlbl_prx}, a two-dimensional bipartite lattice with hopping strength varied in different time steps [see Fig. \ref{fig:rlbl_model}(a)].
Its Hamiltonian reads $H(t)=H_{\text{ons}}+H_{\text{hop}}(t)$ with $H_{\text{ons}}=\delta_{\text{AB}}\sum_{i,j}(c_{A,i,j}^{\dagger}c^\pd_{A,i,j}-c_{B,i,j}^{\dagger}c^\pd_{B,i,j})$  and
\begin{eqnarray}
H_{\text{hop}}(t)=J\sum_{i,j}\begin{cases}
(c_{A,i,j}^{\dagger}c^\pd_{B,i,j}+\text{h.c.}),~t\in\frac{T}{5}[0, 1);\\
(c_{A,i,j}^{\dagger}c^\pd_{B,i-1,j+1}+\text{h.c.}),~t\in\frac{T}{5}[1,2);\\
(c_{A,i,j}^{\dagger}c^\pd_{B,i-1,j}+\text{h.c.}),~t\in\frac{T}{5}[2,3);\\
(c_{A,i,j}^{\dagger}c^\pd_{B,i,j-1}+\text{h.c.}),~t\in\frac{T}{5}[3, 4);\\
0,~t\in\frac{T}{5}[4, 5).
\end{cases}
\end{eqnarray}
Here, $t$ is time, $J$ is the hopping strength, and $c_{A,i,j}^\dagger$ ($c^\pd_{B,i,j}$) denotes the creation (annihilation) operator for sublattice $A$ ($B$) in unit cell $(i,j)$ [see Fig.~\ref{fig:rlbl_model}(a)]. 
Setting $\hbar=1$, the dynamics of the system is governed by a Floquet operator $\mathcal{F}=\mathcal{T}e^{-\int_{0}^{T}iH(t)dt}$, with ${\cal T}$ denoting time ordering. The quasienergy spectrum $\varepsilon$ is contained in the fundamental domain $[-\pi/T,\pi/T)$ and can be obtained from the eigenvalue equation $\text{det}[\mathcal{F}-e^{-i\varepsilon T}]=0$.

In a finite-sized geometry, when switching off the sublattice potential ($\delta_{\text{AB}}=0$), there are two limits in which the Floquet operator takes a particularly simple form. 
For the trivial limit with $JT=0$, we have $\mathcal{F}=1$ and no particles can propagate.
In the AFTI limit with $JT=5\pi/2$ (also known as the resonant driving point \cite{Fulga_fti}), 
all bulk states come back to their original sites after one driving cycle, forming two degenerate, dispersionless Floquet bulk bands at $\varepsilon=0$.
Even if states do not propagate throughout the bulk, an extra quantized conducting channel is formed at the edge, allowing particles to propagate unidirectionally. 
This is the chiral edge mode of the AFTI, which winds in quasienergy from $-\pi/T$ to $\pi/T$. 

At resonant driving, we draw a connection between Floquet and non-Hermitian topology by using a duality that identifies time-evolution operators with non-Hermitian Hamiltonians.
The latter has been used to study the bulk spectral properties of non-Hermitian systems \cite{Floquet_Hamiltonian2, Huang_prb_duality}, and also for the purpose of topological classification \cite{kawabata_prx, gong_prx}.
Here, instead, we focus on its boundary manifestations.
We obtain a unitary, static tight-binding model by treating the real-space Floquet operator as a Hamiltonian:
\begin{eqnarray}\label{eq:FtoH}
H_{\text{u}}=\mathcal{F}.
\end{eqnarray}
Mathematically, this means that we represent the Floquet operator of the finite-sized sytem as a matrix in the real-space basis corresponding to the site positions of the lattice in Fig.~\ref{fig:rlbl_model}(a).
Thus, in matrix representation,
$[H_{\text{u}}]_{jk} \equiv {\cal F}_{jk} = \bra{{\bf x}_k} {\cal F} \ket{{\bf x}_j}$, with $\ket{{\bf x}_j}$ the position ket of site $j$.
The diagonal entries, ${\cal F}_{jj}$, become (possibly complex-valued) on-site terms, whereas the off-diagonal terms, ${\cal F}_{jk}$ with $j\neq k$, are (possibly nonreciprocal) hoppings.

At $\delta_{\text{AB}}=0$ and $JT=5\pi/2$, the real-space structure of $H_\text{u}$ is shown in Fig.~\ref{fig:rlbl_model}(b).
Its bulk contains decoupled sites with unit on-site potentials, ${\cal F}_{jj}=1$, consistent with the existence of dispersionless bulk bands at energy $E=1$ (meaning quasienergy $\varepsilon=0$ in Floquet language).
On the boundary, however, the unidirectional propagation of particles leads to one-way hoppings, ${\cal F}_{jk}=1$, where $j$ and $k$ correspond to sites connected by an arrow in Fig.~\ref{fig:rlbl_model}(b).
The AFTI boundary is identical to a maximally-nonreciprocal Hatano-Nelson chain with periodic boundary conditions (PBC) \cite{HN_prl}.
As such, it will show the same phenomenology as the Hatano-Nelson chain: all states become localized at one end when changing from periodic to open boundary conditions.
We achieve the latter by
removing one hopping from the chain (setting the off-diagonal term corresponding to the arrow in the shaded ellipse to ${\cal F}_{jk}=0$).
The propagation of the Floquet chiral edge modes stops, leading to the formation of a non-Hermitian skin effect.

Note that this behavior is different from the recently-introduced `hybrid skin-topological modes' of Refs.~\cite{hybrid_skin_1, hybrid_skin_2, Linhu_hybrid, Lee_hybrid}, which are generated by adding gain and loss to either static or periodically-driven systems.
Here, the 1D non-Hermitian topology is intrinsic to the AFTI phase, it is the boundary manifestation of a 2D AFTI bulk, and removing one hopping simply serves to change the Hatano-Nelson chain from PBC to OBC.

\section{Robustness of the non-Hermitian skin effect}
\label{sec:skin_effect}
The emergence of 1D non-Hermitian topology at the boundary of a 2D AFTI bulk is most apparent at resonant driving, due to the simple form of the Floquet operator.
Away from this point, $H_{\text{u}}$ contains longer-range hopping terms (up to 4$^{\rm th}$-nearest neighbor, see Supplemental Material -- SM \cite{SupMat}), such that a direct, visual identification of the boundary Hatano-Nelson chain is no longer possible.
Nevertheless, as we show in the following, the skin effect formed by removing a boundary hopping remains robust, provided that the system is in an AFTI phase.

We study the phase diagram of the Floquet system numerically.
We use the kwant package \cite{Groth_2014}, provide details on the numerical simulations in the SM \cite{SupMat}, and share our code on Zenodo \cite{zenodo}.
When $\delta_{\rm AB}=0$, the AFTI and trivial phase are separated by a topological phase transition at $JT=5\pi/4$, at which the bulk gap around $\varepsilon=\pi/T$ closes at the $\Gamma$ point of the Brillouin zone.
At $JT=15\pi/4$, a second phase transition occurs at $\varepsilon=\pi/T$ and drives the system back to a trivial phase. 
No gap closing and reopening occurs at $\varepsilon=0$ for any value of $JT$,
provided that $\delta_{\text{AB}}=0$.

If $\delta_{\text{AB}}\neq 0$, however, the $\varepsilon=0$ gap will open,
but with a gap size that vanishes as $\delta_{\text{AB}}\to 0$, see Fig.~\ref{fig:rlbl_model}(c). 
Due to this extra gap, Chern insulating (CI) phases appear, which separate the trivial from the AFTI phase, as shown in Fig.~\ref{fig:rlbl_model}(d). 
Thus, the CI phase on the left side of the phase diagram in Fig.~\ref{fig:rlbl_model}(d) transforms into an AFTI due to the closing of the $\varepsilon=0$ gap at $\delta_{\text{AB}}=0$.
In contrast, the CI on the right hand side of the phase diagram is transformed into a trivial phase (see SM for bandstructure plots detailing this process \cite{SupMat}).
Formally, the CI phases persist for arbitrarily small values of $\delta_{\text{AB}}$, provided that $\delta_{\text{AB}} > 0$, though in practice their observation will be hindered by the smallness of the $\varepsilon=0$ gap, which leads to significant finite-size effects.

Throughout the AFTI phase, even though the longer-range hoppings in $H_{\rm u}$ couple the edge and the bulk sites, there exists a large density accumulation at the site next to the defect.
To show this, we define the real space probability density
\begin{equation}\label{eq:rho}
\rho_{{\bf x}_j} = \sum_n \braket{{\bf x}_j | \psi_n},
\end{equation}
where the sum runs over all of the right eigenvectors of the Floquet operator $\ket{\psi_n}$.
The color scale of Fig.~\ref{fig:rlbl_model}(d) corresponds to the real-space probability density on the site just above the cut hopping, $\rho_{{\bf x}_c}$.

Notably, there is a region of large density also inside the left-most CI phase, the one which evolves into an AFTI as $\delta_{\text{AB}}$ goes to 0.
In contrast, no density accumulation occurs in the right-most CI, a phase which becomes trivial for $\delta_{\text{AB}}=0$.
We attribute this additional `shoulder' of large $\rho_c$ to the proximity of the $\delta_{\text{AB}}=0$ AFTI phase.
The skin effect is otherwise absent from the CI phases.

\begin{figure}
	\centering
	\includegraphics[width=\linewidth]{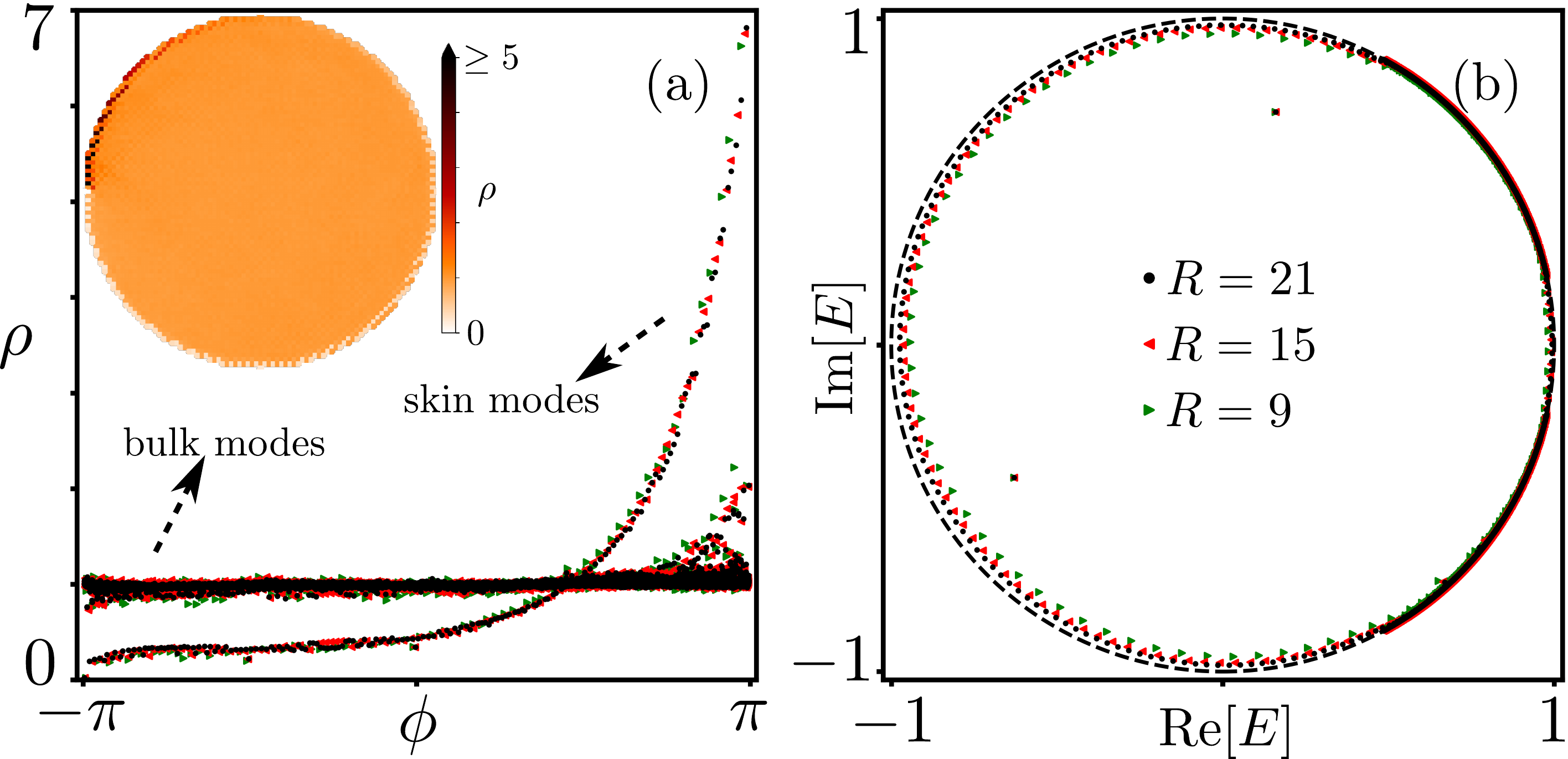}
	\caption{The left panel is the real-space probability density as a function of polar angle $\phi$ in the disk geometry with different radius $R$. 
	The inset is the same result in Cartesian coordinate representation with $R=21$.
	Here, the polar coordinate origin is fixed to make the defect at $\phi=\pm\pi$.
	The right panel is the corresponding energy spectrum. 
	All plots are with $JT=2.7\pi$ and $\delta_{\text{AB}}T=0.75\pi$.
	See SM for numerical details \cite{SupMat}.
		\label{fig:skin_effect}
	}
\end{figure}

We investigate the properties of the skin effect in a disk geometry, going to polar coordinates $r$ and $\phi$.
The real-space probability density corresponding to each of the sites [Eq.~\eqref{eq:rho}] is plotted as a function of the angular position $\phi$ in Fig.~\ref{fig:skin_effect}(a), and shows two distinct behaviors.
In the bulk of the disk (see also the inset), the probability density is uniform, as expected in Floquet systems.
On the boundary, however, it is peaked at the defect position, showing a profile that gradually decays as a function of $\phi$, while remaining pinned to the edge of the disk.
Interestingly, the probability density curves obtained for increasing disk radius $R$ overlap in Fig.~\ref{fig:skin_effect}(a), which means the skin effect of the chiral modes is a scale-invariant phenomenon.
This behavior has been dubbed the `critical skin effect' in Ref.~\cite{Li2020}, and has been predicted to occur when two or more non-Hermitian subsystems are coupled to each other.
In our case, we conjecture that it results from the nonzero coupling between the edge and the bulk away from resonant driving.
As a final check of the critical nature of the skin effect, we confirm in Fig.~\ref{fig:skin_effect}(b) one of its predicted hallmark features: when the system size is increased, its eigenvalues approach the ones of the infinite system (here, the unit circle)~\cite{Yokomizo_critical}.

\section{Real-space topological invariant}
\label{sec:real_space_invariant}
The robustness of the skin effect suggests that it is topologically protected.
This is clearly the case at resonant driving, where the boundary of the real-space Floquet operator is decoupled from the bulk, forming a purely 1D Hatano-Nelson Hamiltonian with PBC.
For the latter, bulk-boundary correspondence allows to predict the accumulation of edge modes under OBC by computing the winding number of the bulk spectrum \cite{topological_origin}.
Away from resonant driving, however, the nonzero hoppings between bulk and edge sites of the 2D Floquet operator obscure the distinction between PBC and OBC.

To overcome this difficulty, we instead introduce a real-space topological invariant which predicts the skin effect of the chiral edge modes starting from the full, 2D Floquet operator.
Our first ingredient is the by now well-established correspondence between Hermitian and non-Hermitian topological phases \cite{topological_origin,Borgnia_prl}. 
According to it, if $H_{\text{nH}}$ is a 1D non-Hermitian Hamiltonian that is topologically equivalent to the Hatano-Nelson model, then 
\begin{eqnarray}\label{eq:Hdoubling}
\tilde{H}_{\text{H}}=\begin{pmatrix}
0&H_{\text{nH}}-E_{\text{b}}\\
H_{\text{nH}}^{\dagger}-E^{*}_{\text{b}}&0
\end{pmatrix}
\end{eqnarray}
is a Hermitian Hamiltonian that is topologically equivalent to a Su-Schrieffer-Heeger (SSH) chain \cite{SSH_MODEL}.
More precisely, the winding number of the spectrum of $H_{\text{nH}}$ around the base point $E_{\text{b}}$ is identical to the SSH topological invariant of $\tilde{H}_{\text{H}}$.

Our second ingredient is based on the fact that, for a 1D finite SSH chain with OBC, previous work has shown that the topological invariant can be determined from a quantity known as the `spectral localizer'~\cite{LORING2015383, Loring2017, Cerjan_prb, Cheng_arxiv}.
This is a Hermitian matrix which directly measures the topological zero energy modes of $\tilde{H}_{\text{H}}$ with OBC, taking the form
\begin{eqnarray}
L=[X+i\tilde{H}_{\text{H}}]Q, \label{eq: sig}
\end{eqnarray}
where $X$ is the position operator and $Q=\text{Diag}[\mathbb{I},-\mathbb{I}]$ is the chiral symmetry operator, with $\mathbb{I}$ an identity matrix of the same size as $H_{\text{nH}}$ (see SM \cite{SupMat}).
The real-space topological invariant is defined as the signature of the spectral localizer (denoted $\text{sig}[L]$), meaning the number of positive eigenvalues of $L$ minus the number of negative eigenvalues.

\begin{figure}
	\centering
	\includegraphics[width=\linewidth]{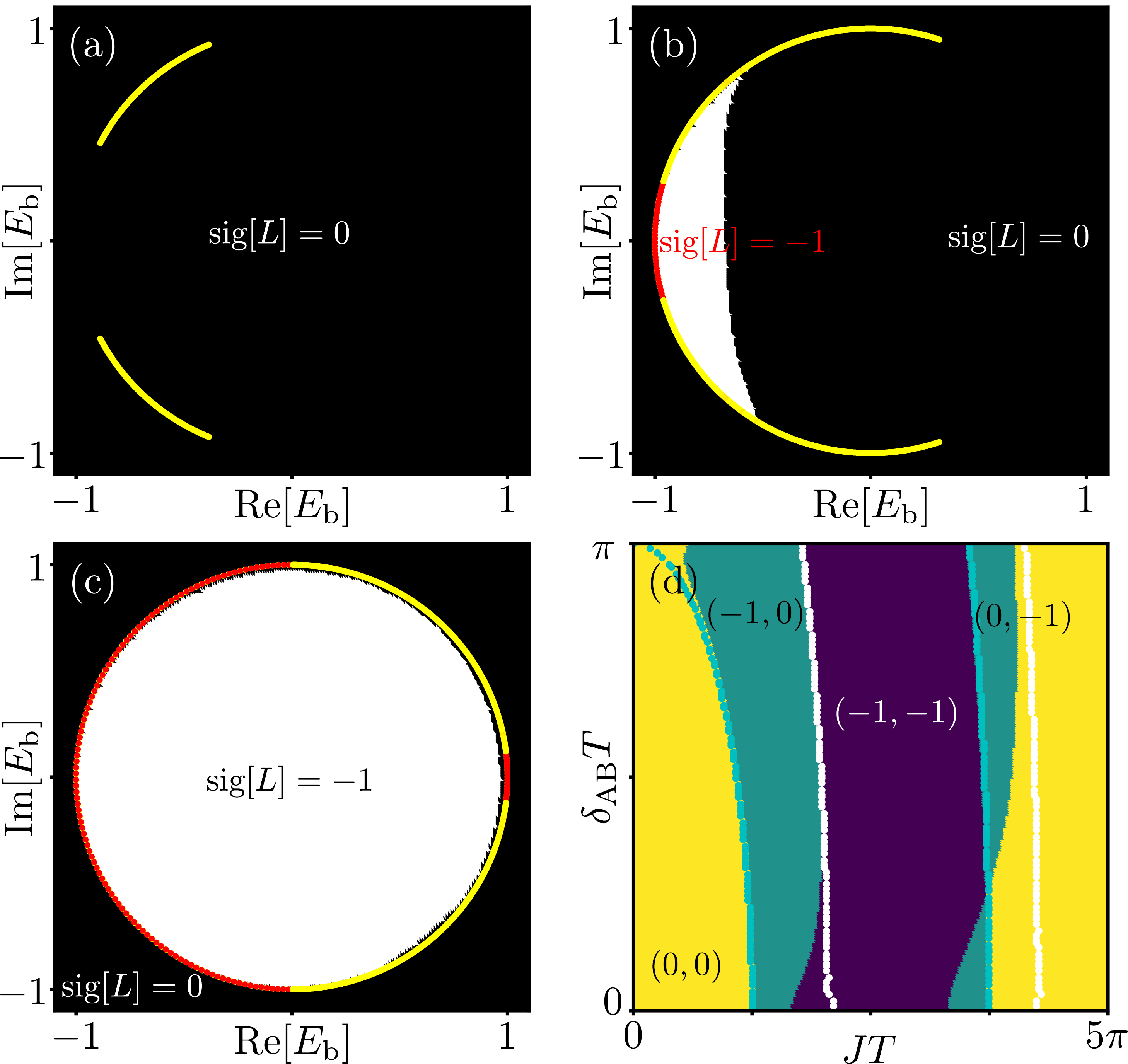}
	\caption{Signature as a function of $\text{Re}[E_{\text{b}}]$ and $\text{Im}[E_{\text{b}}]$ for (a) $JT=0.5\pi$ (trivial), (b) $JT=\pi$ (CI), and (c) $JT=2\pi$ (AFTI) with $\delta_{\rm AB}T=0.75\pi$, respectively. 
	The yellow and red lines represent bulk and edge states of $H_{\text{u}}$, respectively. 
	(d) The phase diagram shows the pair of signatures at base energies $-0.9$ and $0.9$ as a function of $\delta_{\rm AB}$ and $J$. 
	The values $(0, 0)$, $(-1, -1)$, and $(-1, 0)$ or $(0, -1)$ are shown as yellow, dark blue, and green, respectively.
	See SM for numerical details \cite{SupMat}.
		\label{fig:topological_invariant}
	}
\end{figure}

We use Eq.~\eqref{eq: sig} to study the 2D system in the presence of a boundary defect by replacing $H_{\rm nH}$ with $H_{\rm u}$ in the off-diagonal blocks of Eq.~\eqref{eq:Hdoubling}.
The effective Hatano-Nelson model is now no longer positioned in a 1D space with position operator $X$, but at boundary of a 2D disk.
Therefore, we replace $X$ in Eq.~\eqref{eq: sig} with the angular position operator in polar coordinates, $\Phi$ (see SM \cite{SupMat}).

The signature vanishes in the trivial phase at any base energy, consistent with the absence of a skin effect [see Fig.~\ref{fig:topological_invariant}(a)]. 
In the AFTI phase, on the other hand, the skin effect is present and $\text{sig}[L]=-1$ everywhere inside of the unit disk, as shown in Fig.~\ref{fig:topological_invariant}(c).
As the disk becomes larger and larger, the density of eigenvalues increases, eventually covering the full unit circle in the thermodynamic limit [see Fig.~\ref{fig:skin_effect}(b)].
Thus, in non-Hermitian language, there exists a nontrivial point gap at $|E_{\rm b}|<1$, the region of the complex plane enclosed by the system spectrum, as expected in non-Hermitian topological phases hosting the skin effect~\cite{kawabata_prx, gong_prx}.
Note that this is a nontrivial invariant of the full, 2D Floquet operator, $H_{\rm u}={\cal F}$, which contains both bulk as well as edge states, and as such attests to the presence of higher-order topology.

In the CI phase [Fig.~\ref{fig:topological_invariant}(b)], however, no point gap is formed and we observe a gradual transition in which a sliding domain wall appears between $\text{sig}[L]=-1$ and 0 regions of the complex plane, interpolating between the trivial and AFTI results of panels (a) and (c), as shown in the SM \cite{SupMat}.
The position of this domain wall depends on the size of the gaps at $E_{\rm b}=+1$ and $-1$ ($\varepsilon=0$ and $\pi/T$ in Floquet language), as well as on the number of hoppings removed from the system to produce the defect.
The larger the bulk gap at either of these two (quasi-) energies, or the more hoppings are removed from $H_{\rm u}$, the smaller the nontrivial region of $E_{\rm b}$.
This suggests that the latter is a finite-size effect, consistent with our expectation that the skin effect should not be protected in the absence of a point gap.

These findings provide an indication that the non-Hermitian point gap topology determines whether the chiral edge modes stop (forming a skin effect, as in the AFTI), or continue propagating around the cut hopping (as in the CI).
For a more complete picture, 
we present a phase diagram of the topological invariant at two opposite base energies in Fig.~\ref{fig:topological_invariant}(d). 
It shows that ${\rm sig}[L]$ provides a good description of both the trivial phase and the AFTI phase for a wide range of onsite potentials $\delta_{\rm AB}$ and hopping strengths $J$, while suffering from finite size effects in the CI regions and close to $\delta_{\rm AB}=0$.

\section{Conclusion and outlook}
\label{sec:conclusion}
In Refs.~\cite{Floquet_Hamiltonian1, H_ckendorf_2020}, it was shown that adding non-Hermitian perturbations to the boundary of a periodically-driven system can enhance the robustness of AFTI edge modes by decoupling them from the bulk.
Here, instead, we have shown that non-Hermitian topology is an intrinsic boundary manifestation of the AFTI bulk: it occurs without the need for any external perturbation.
As such, a vanishingly small perturbation, removing one hopping from an arbitrarily large Floquet disk, causes the edge modes to stop altogether.
They accumulate at the defect position instead, forming a non-Hermitian skin effect.

This effect represents an alternate, `mixed' form of higher-order topology, where the boundary of a two-dimensional system characterized by Floquet topology realizes one-dimensional non-Hermitian topology.
The latter appears when Floquet edge modes wind in quasienergy, it is robust against changing parameter values, and it is general; we have checked that it occurs for a variety of different AFTIs, as well as network models. 
Along this direction, it would be interesting to see how this skin effect can form in an extrinic Floquet setting, the quantum walks~\cite{extrinic_quantum_walk}.

We have shown that this effect is topological by adapting a tool that has so far been used mainly for Hermitian systems: the spectral localizer.
Finding other methods of studying the robustness of this type of higher-order topology, as well as exploring the number of dimensions, symmetry classes, and space groups in which it may occur, are interesting directions for future work.

On a more practical level, our work suggests a potentially simpler way of obtaining the non-Hermitian skin effect, by starting from periodically-driven topological phases.
The latter have matured over the last decade, and can now be reliably demonstrated in a variety of experimental platforms, ranging from photonic crystals and coupled ring resonators to acoustic systems and ultracold atoms~\cite{afti_ex0, afti_ex1, afti_ex2, afti_ex3, afti_ex4, afti_ex5, afti_ex6, afti_ex7, afti_ex8, afti_ex9, afti_ex10}.
Physically, changing the boundary conditions of the non-Hermitian chain by cutting a hopping from the real space Floquet operator corresponds to introducing loss to a single point on the edge of the system.
In contrast, the emerging experiments on non-Hermitian topology mainly focus on gain and loss (or on nonreciprocity) which is tailored such as to occur in a specific pattern throughout the entire bulk of the system~\cite{nh_ex1, nh_ex2, nh_ex3, nh_ex4, nh_ex5, nh_ex6, nh_ex7, nh_ex8, nh_ex9, nh_ex10, nh_ex11, nh_ex12, nh_ex13}.

\section{Acknowledgements}
We thank Ulrike Nitzsche for technical assistance and Viktor K\"{o}nye for helpful discussions. We acknowledge financial support from the DFG through the W{\"u}rzburg-Dresden Cluster of Excellence on Complexity and Topology in Quantum Matter – ct.qmat (EXC 2147, project-id 390858490).

\bibliography{reference}

\end{document}